\documentstyle[amssymb,twocolumn,prl,aps,epsfig]{revtex}

\begin{document}
\draft

\twocolumn[\hsize\textwidth\columnwidth\hsize\csname
@twocolumnfalse\endcsname

\title{Spin-Peierls transition with strong structural
fluctuations in the vanadium oxide
VOSb$_{2}$O$_{4}$
}

\author{V.A. Pashchenko$^{1,2}$, A. Sulpice$^{3}$, F. Mila$^{4}$,
P. Millet$^{5}$, A. Stepanov$^{6}$, and P. Wyder$^{1}$}

\address{$^{1}$ Grenoble High Magnetic Field Laboratory, MPI-FKF and CNRS,
BP 166, 38042 Grenoble Cedex 9, France \\
$^{2}$ B.Verkin Institute for Low Temperature Physics and Engineering,
NASU, 310164 Kharkov, Ukraine\\
$^{3}$ Centre de Recherches sur les Tr\`es Basses Temp\'erautres, CNRS,
BP 166, 38042 Grenoble Cedex 9, France\\
$^{4}$ Institut de Physique Th\'eorique, Universit\'e de Lausanne,
CH-1015 Lausanne, Switzerland\\
$^{5}$ Centre d'Elaboration des Mat\'eriaux et d'Etudes Structurales, CNRS,
BP 4347, 31055 Toulouse Cedex 4, France\\
$^{6}$ Laboratoire Mat\'eriaux et Micro\'electronique de Provence,
Universit\'{e} d'Aix-Marseille III, CNRS,\\
Facult\'{e} des Sciences de Saint-J\'{e}r\^ome, C-151, 13397 Marseille Cedex 20, France\\}

\date{\today}
\maketitle

\begin{abstract}
We report on the magnetic susceptibility and electron spin resonance
measurements on polycrystalline samples of the vanadium oxide
VOSb$_{2}$O$_{4}$, a quasi-one dimensional S=1/2 Heisenberg system.
We show that the susceptibility vanishes at zero temperature, as in
a gapped system, and we argue that this is due to a spin-Peierls transition
with strong structural fluctuations.

\end{abstract}

\vskip2pc] \narrowtext

Although the study of the spin-Peierls (SP) transition in $S=1/2$
antiferromagnetic (AF) Heisenberg chains has started a long time ago with
the discovery of the first SP transition in the organic system TTFCuBDT in
1975\cite{bray,jacobs}, a major breakthrough in the field was the discovery
in 1993 by Hase et al.\cite{hase} of the first inorganic system exhibiting a
SP transition, namely CuGeO$_3$. The possibility to grow large single
crystals has led to a very intensive experimental activity, and the
understanding of the properties of such systems, in particular in strong
magnetic fields, has been dramatically improved.

However CuGeO$_3$ is representative of only one class of spin-Peierls
systems, namely systems in which structural fluctuations are to a certain
extent negligible. In such systems, the dimerization of the lattice is very
brutal, and the susceptibility exhibits a characteristic cusp at the
transition temperature. The irrelevance of structural fluctuations in a 1D
system is very surprising, and the first theories actually predicted a
strongly fluctuating regime above the transition\cite{lee}. This discrepancy
was resolved by Cross and Fisher\cite{cross}, who showed that an appropriate
treatment of 3D phonons can lead to a significant suppression of
fluctuations.

The study of fluctuations in spin-Peierls systems has recently restarted
however with the careful analysis of the spin-Peierls transition in the
organic system (BCPTTF)$_2$X by Dumoulin et al.\cite{dumoulin} in 1996 who
convincingly showed the presence of very strong structural fluctuations
above the spin-Peierls transition. Judging from the impact of CuGeO$_3$ on
the field, the search for inorganic systems with similar properties is a
real challenge. However the inorganic spin 1/2 chains synthetized so far do
not seem to fill this gap. Most of them just do not show any sign of a SP
instability, like Sr$_2$CuO$_3$\cite{ami} or MgVO$_3$\cite{choukroun}, while
the transition observed in NaV$_2$O$_5$\cite{isobe} is very abrupt and is
now believed to involve charge degrees of freedom as well.

In this paper we report on the magnetic properties of a vanadium oxide, VOSb$%
_{2}$O$_{4}$, which we believe is the first example of an inorganic system
that undergoes a SP transition with very strong fluctuation effects. This
system is made of almost isolated chains of VO$_{5}$ pyramids. According to
Darriet, Bovin and Galy \cite{Darriet} VOSb$_{2}$O$_{4}$ crystallizes in the
monoclinic system, space group $C2/c$, with the unit cell dimensions $a$%
=18.03 \AA , $b$=4.800 \AA , $c$=5.497 \AA , $\beta $=94.58${^{\circ }}$
(Z=4). The vanadium atoms are fivefold coordinated in a slightly distorted
square pyramids with one characteristic short vanadyl bond V-O close to 1.59
\AA {} towards its apex and 2$\times $2 longer bonds at 1.91 and 2.04 \AA {}
with the oxygens of the square base. Along the [001] direction the apices of
the VO$_{5}$ pyramids alternately point up and down relative to a plane of
the square base (see Fig. 1(a)). The smallest in-chain V-V distance is
approximately 3.01 \AA . The distances between the chains are 4.80 and 18.03
\AA {} for the [010] and the [100] direction, respectively. Thus
from a magnetic point of view, the VOSb$_{2}$O$_{4}$ structure can be viewed
as infinite isolated chains of V$^{4+}$ ions running along the [001]
direction. The antimony atoms exhibit the typical one-sided threefold
coordination of the oxygen atoms having a stereoactive lone pair $E\;$ \cite
{Darriet} (see Fig. 1(b)).

Polycrystalline samples of VOSb$_{2}$O$_{4}$ \ having a light-green color
were synthesized by solid-state reaction \cite{Darriet}. ESR X-band spectra
were collected using a Bruker ESP300 spectrometer equipped with a standard TE%
$_{102}$ cavity and a continuous helium flow cryostat that allows
temperature scans between 4 and 300 K. The temperature and the field
variation of the magnetization was measured with a Quantum Design SQUID
magnetometer from 300 to 1.8 K in fields up to 4 T.

\begin{figure}[ph]
\mbox{(a)}
\begin{center}
\includegraphics*[width=65mm,angle=0]{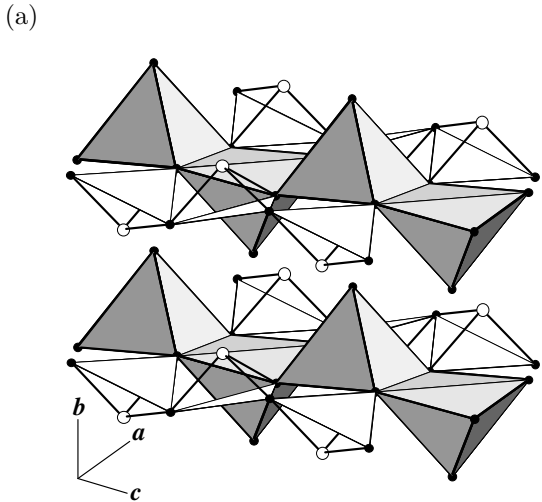}
\end{center}
\mbox{(b)}
\begin{center}
\includegraphics*[width=65mm,angle=0]{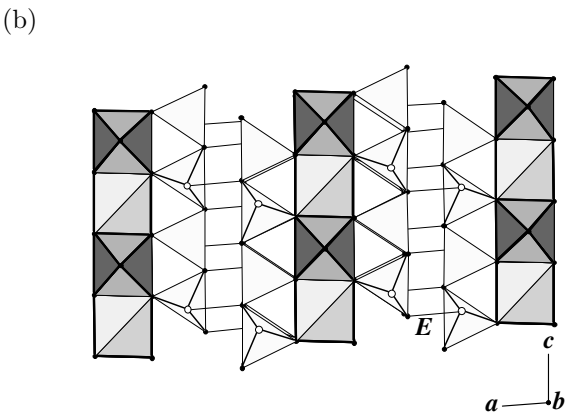}
\end{center}
\caption{(a) Polyhedral representation of the VOSb$_{2}$O$_{4}$ structure
illustrating the infinite isolated chains of VO$_{5}$ pyramids. (b) the
($ac$) plan projection of the crystal structure of VOSb$_{2}$O$_{4}.$}
\label{fig1}
\end{figure}

The temperature dependence of the magnetic susceptibility $\chi _{raw}(T)$
of a 100 mg polycrystalline sample of VOSb$_{2}$O$_{4}$ in a field of 2 T is
shown in Fig. 2. Below room temperature, when the temperature is lowered,
$\chi (T)$ passes through a broad maximum at $T_{\max }\approx 160$ K, which
is typical of a $S=1/2$ Heisenberg chain with $J/k\simeq 250$ K. However, on
further cooling the sample, there is a rapid decrease which starts around
$40$ K, and the slope exhibits a clear maximum at $T_{sp}=13$ K.
Finally, there is a minimum at $10$ K followed by an increase of the
susceptibility indicating the presence of magnetic impurities. The drop that
starts around $40$ K is reminiscent of a SP transition, but there is a
dramatic difference with CuGeO$_{3}$: there is no cusp in the susceptibility
around the temperature where it becomes much smaller than the Bonner-Fisher
prediction. On the contrary, the susceptibility drops rapidly but smoothly
below $35$ K, very much like in (BCPTTF)$_{2}$X.

\begin{figure}[ph]
\begin{center}
\includegraphics*[width=90mm,angle=0]{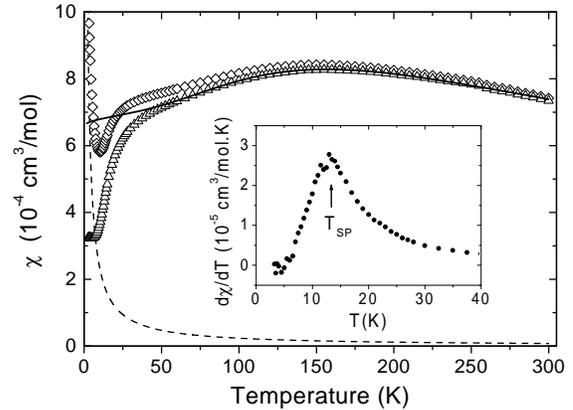}
\end{center}
\caption{The temperature dependence of the magnetic susceptibility of a 100
mg VOSb$_{2}$O$_{4}$ polycrystalline sample $\protect\chi _{raw}(T)$ at $H=$
2 T (open diamonds). Open triangles, dashed line and solid line represent
the corrected for the impurity contribution VOSb$_{2}$O$_{4}$
susceptibility $\protect\chi _{cor}(T)$, the impurity contribution to
$\protect\chi _{raw}(T)$ and the best fit to the $\protect\chi (T)$ of $S=1/2$
HAF according to Eq. (50), Ref. 18 ( $J/k=245\pm 5$ K, $\protect\chi %
_{0}=0.00021$ cm$^{3}$/mol, $g=1.66$), respectively. The insert details low
temperature behavior of $d\protect\chi _{cor}(T)/dT$.}
\label{fig2}
\end{figure}

Before we can start discussing the physical origin of this unusual
behaviour, the first thing we must check is whether the susceptibility
indeed goes to zero at zero temperature, as in a SP transition. Let us first
study in more details the impurity contribution. To characterize more
quantitatively this impurity contribution $\chi _{imp}(T)$ to $\chi
_{raw}(T) $,\ particularly with an idea to separate it from the intrinsic
susceptibility of the VOSb$_{2}$O$_{4}$ phase, which we will call $\chi
_{cor}(T),$we have carried out magnetization measurements vs. $H$ at
various fixed temperatures from $80$ K to $1.8$ K. The results are shown in
Fig. 3. An important information about $\chi _{imp}(T)$ is contained in the
low temperature nonlinear dependence of $M_{imp}(H,T)$ when $\mu H>kT$.
Therefore we have to examine the data in Fig. 3 using the following equation $%
M(H,T)=M_{imp}(H,T)+\chi _{cor}(T)\cdot H,$ where $M_{imp}(H,T)$ is
expressed in a standard way as $\ M_{imp}(H,T)$=$p_{imp}N_{A}g\mu _{B}S\cdot
B_{S}(g\mu _{B}SH/kT)$, $B_{S}$ is the Brillioun function, $S$ \ is an
impurity spin value and $p_{imp}$ defines the relative impurity
concentration. The results of a fit of the experimental data with this
equation are shown in Fig. 3 as solid lines. The following parameters $S=1/2$%
, $p_{imp}$=0.00573(5) were extracted ( $g$=1.975 was fixed in the fit
procedure) together with the AF Curie-Weiss constant $\theta \approx 0.6$ K
obtained from the low-$T$ dependence of $\chi _{raw}^{-1}(T)$. We are now in a
position to correct $\chi _{raw}(T)$ for the impurity contribution. In Fig. 2
$\chi _{imp}(T)=\,$$p_{imp}C/(T+\theta )$ ($C=0.366$ cm$^{3}\cdot$K/mol)
is plotted (dashed line) along with the $\chi _{cor}(T)=$$\chi
_{raw}(T)-\chi _{imp}(T)$. This behaviour is consistent with a zero
contribution to the spin susceptibility $\chi _{spin}(T)=\chi
_{cor}(T)-\chi _{0}$ at zero temperature if the sum of the diamagnetic and
Van Vleck contributions $\chi _{0}$ is equal to $\chi _{cor}(T=0)$. While
the diamagnetic contribution can be estimated from standard tables ($\chi
_{dia}\simeq -1.01\times 10^{-4}$cm$^{3}/$mol \cite{Selwood}), an unbiased
estimate of the Van Vleck susceptibility would require susceptibility data
at temperatures much larger than the typical exchange integrals, a regime
which is not accessible.

\begin{figure}[ph]
\begin{center}
\includegraphics*[width=85mm,angle=0]{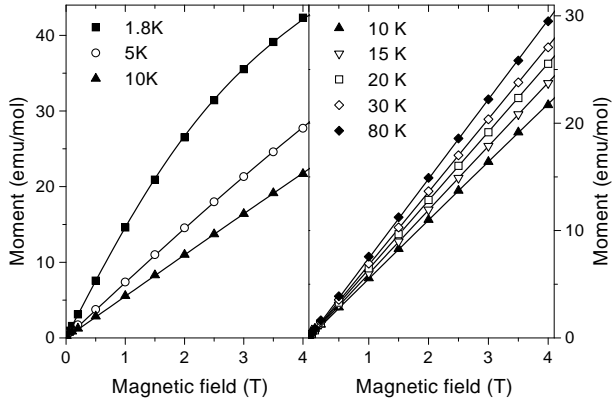}
\end{center}
\caption{The magnetization of VOSb$_{2}$O$_{4}$ vs. magnetic field at
different temperatures. The symbols are experimental data, the solid lines
are the fit results according to the equation of $M(H,T)$ discussed in the
text.}
\label{fig3}
\end{figure}

To go around this difficulty, we have performed extensive ESR measurements.
A representative series of X-band ESR spectra recorded from $320$ to $7.3$ K$%
\ $on a polycrystalline sample (12 mg) is presented in Fig. 4(a). We note the
axial symmetry of the obtained spectra, especially apparent at $T=20$ K,
reflecting the axial symmetry of the crystal field acting on V$^{4+}$ ions
in the fivefold pyramidal environment. The computer spectra simulations over
the temperature range 13 - 320 K give two $T$-independent $g$-factors:
$g_{\perp }$=1.978, $g_{\parallel }$=1.930 with the average value of
$1.962\pm 0.002$ already reported for low-dimensional vanadates \cite
{choukroun,yamada}. At low $T$ it is found that the measured spectra
contain an additional ESR signal. The intensity of this additional signal
roughly follows Curie law and the average $g-$factor is found to be
$1.975\pm 0.005$. We ascribe this signal to the magnetic impurities which are
responsible for the steep increase of the magnetic susceptibility at low
temperatures (see above).

To extract information from these ESR spectra, we have proceeded in the
following way. Since the ESR is insensitive to the diamagnetic and Van Vleck
contributions to the susceptibility, we are able, by double integration of
the ESR spectra, to reconstruct the sum $\chi _{spin}(T)+$$\chi _{imp}(T)$
and then using an appropriate procedure for the substraction of $\chi
_{imp}(T)$ to restore the $T$-dependence of $\chi _{spin}(T)$.
\begin{figure}[ph]
\mbox{(a)}
\begin{center}
\includegraphics*[width=60mm,angle=0]{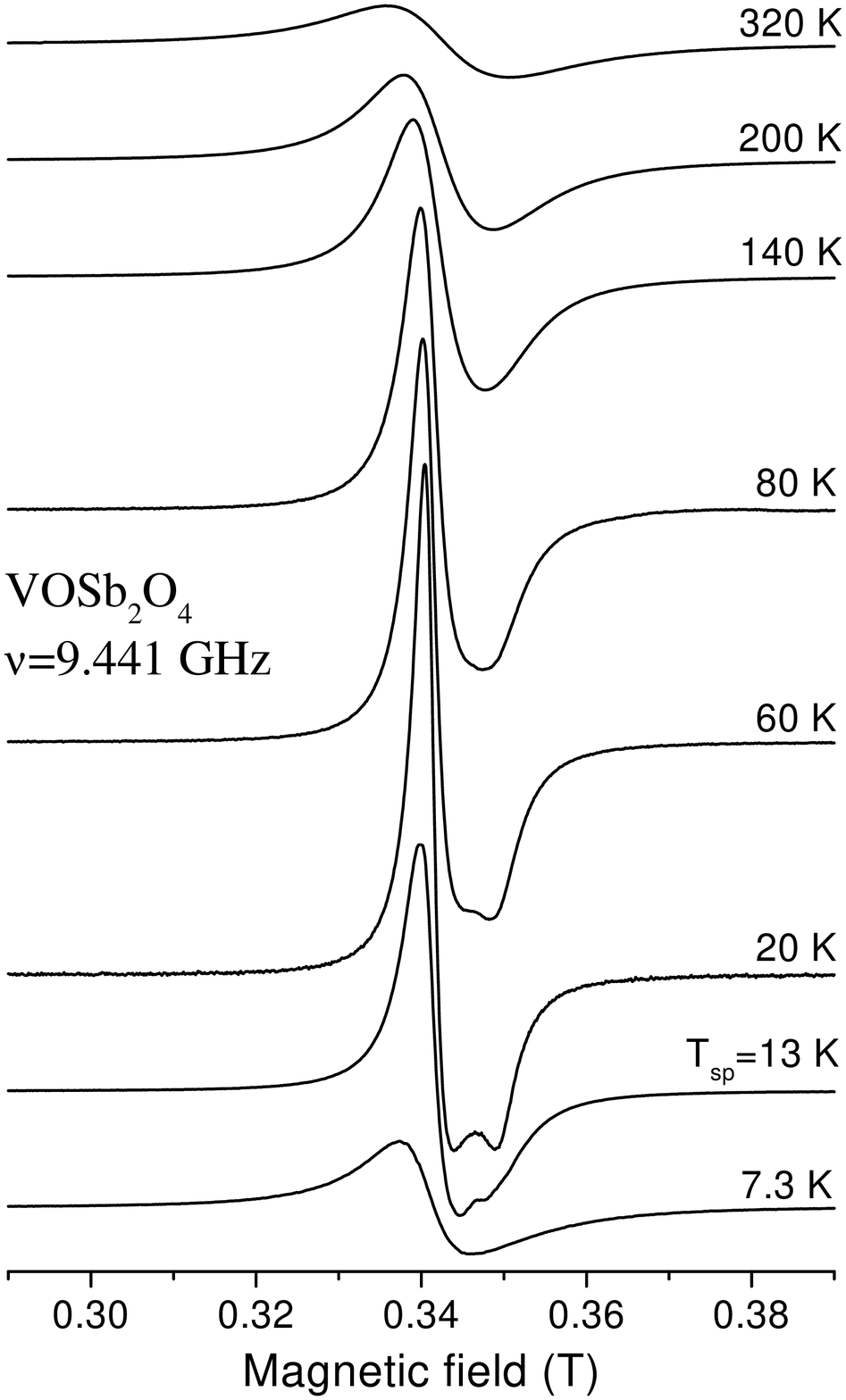}
\end{center}
\mbox{(b)}
\begin{center}
\includegraphics*[width=75mm,angle=0]{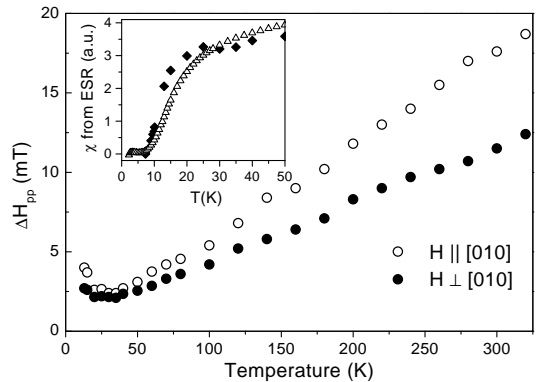}
\end{center}
\caption{(a) The temperature evolution of ESR spectra in the VOSb$_{2}$O$_{4}$
polycrystalline sample from 320 to 7.3 K; (b) the temperature dependence of
the peak-to-peak linewidth $\Delta H_{pp}$ in VOSb$_{2}$O$_{4}$ extracted
from computer simulation of experimental spectra. The insert shows the
double integrated ESR spectra intensity $\protect\chi _{spin}^{ESR}(T)$
(solid diamonds). $\protect\chi _{spin}(T)=\protect\chi _{cor}(T)-\protect%
\chi _{0}$ deduced from Fig. 2 is also shown scaled with $\protect\chi %
_{spin}^{ESR}(T)$ so, that $\protect\chi _{spin}(T_{\max })=\protect\chi
_{spin}^{ESR}(T_{\max }).$}
\label{fig4}
\end{figure}
In the case of VOSb$_{2}$O$_{4}$ the substraction of $\chi _{imp}(T)$ is a rather
tedious but unambiguous procedure because i) the impurity ESR\ spectra have
quite different line parameters (such as the linewidth, g-factor and the
temperature dependence of spectra intensity) as compared to the main
spectra; ii) from the magnetization measurements we know the $T-$evolution
of $\chi _{imp}(T)$ so, we can use this information to check the correctness
of the substraction at each $T$. We drop out the technical details of this
procedure and postpone them to our forthcoming paper. The $\chi
_{spin}^{ESR}(T)$ data as extracted from ESR are given in an inset of
Fig. 4(b). It is clearly seen that the spin magnetic susceptibility of VOSb$%
_{2} $O$_{4}$ goes to zero in the limit $T\rightarrow 0$, a result which is
qualitatively apparent from the examination of ESR spectra at 20, 13 and 7.3
K in Fig. 4(a). For example, the 7.3 K ESR\ spectrum is almost for 95\% an
impurity one. The fact that $\chi _{spin}(T\rightarrow 0)$$\thickapprox 0$
clearly evidences that the VOSb$_{2}$O$_{4}$ ground state at low temperature
is a nonmagnetic singlet $S=0$. Note that the temperature dependence is
consistent with that deduced from the susceptibility measurements after
substraction of the impurity contributions and assuming that the Van Vleck
contribution is such that $\chi _{cor}=0$ (see inset of Fig. 4(b)).

Another very useful information is contained in the temperature dependence
of the line width. As seen from Fig. 4(b), the peak-to-peak linewidth $\Delta
H_{pp}$ shows a characteristic V-like temperature dependence (a strong
decrease of $\Delta H$ replaced at $T\thickapprox 13$ K by a rapid increase
of $\Delta H$). Such a behaviour has been previously observed in both SP
materials NaV$_{2}$O$_{5}$ and CuGeO$_{3}$\cite{lohmann}, the minimum
temperature being equal to the SP transition temperature.

Let us now discuss the various possibilities to explain this behaviour.
Assuming that the chains are well isolated magnetically, which is very
reasonable given the geometry, we can think of only two possibilities to
explain a vanishing susceptibility, hence the presence of a spin gap, at
zero temperature: Frustration or dimerization due to a SP transition. Let us
analyze both possibilities.

{\it Frustration:} It is well known that a coupling $J_{2}$ to second
neighbours can lead to a spin gap if its ratio to the first-neighbour
coupling $J_{1}$ is larger than 0.24\cite{majumdar,okamoto}. However the
presence of a significant coupling between second neighbours will not just
open a gap at low temperature, but will modify the temperature dependence of
the susceptibility at high temperature as well. We have thus tried to fit $%
\chi _{spin}(T)$ with a significant value of $J_{2}$. The resulting fit is
very bad, and much worse actually than without $J_{2}$ between 40 K and 300
K. So this possibility seems unlikely. Besides, if we compare with MgVO$_{3}$%
, another quasi-1D vanadium oxide which does not show any anomaly at low
temperature\cite{choukroun}, the chains have the same structure. But the
magnetic measurements performed on MgVO$_{3}$ show no indication whatsoever
of intra-chain frustration. So it seems more plausible that the difference
between the magnetic properties of these systems comes from the interaction
between the chains. In fact, the chains are further apart in VOSb$_{2}$O$%
_{4} $ than in MgVO$_{3}$, especially in the $a$ direction, where most of
the residual coupling is believed to occur in MgVO$_{3}$. So it is not
surprising that typical 1D effects show up in VOSb$_{2}$O$_{4}$ and not in
MgVO$_{3}$.

{\it Dimerization due to SP transition:} In principle, a $S=1/2$ chain is
always unstable towards dimerization, but the transition temperature can be
strongly reduced due to fluctuations of the lattice, especially if the
system is very one-dimensional. In the present case, a good fit of the
high-temperature susceptibility with the susceptibility of the S=1/2 chain
\cite{BonFish,eggert,Johnston} is possible (see Fig. 2), although with an
effective $g$-factor smaller than the actual one measured in ESR. This
discrepancy is actually ubiquitous in V$^{4+}$ vanadates, whose
properties are quite well understood otherwise, and it seems legitimate not
to worry too much about it. It might come from different factors ranging
from a poor determination of the weight due to the absorption of water by
the sample to the presence of some non-magnetic impurity phase.

The next question is whether we do have a SP transition. From the
susceptibility measurements alone, it is not possible to conclude. But if
there is a transition, it seems likely that it does {\it not} take place at
the onset of the drop, like in CuGeO$_3$, but at the temperature where the
derivative of the susceptibility is maximal, i.e. 13 K. This scenario is
actually favoured by ESR measurements since the line-width changes
dramatically at the same temperature. However, clear signatures of the
transitions, like new Bragg peaks or new phonons lines below 13 K, are not
available yet.

If on the contrary the system remains fluctuating with a pseudo-gap down to
zero temperature, as in the Lee-Rice-Anderson theory of the Peierls
transition in metallic systems \cite{lee}, the susceptibility is expected to
decrease smoothly to zero. This would be consistent with our data. The
behaviour of the line-width under such circumstances is not known however,
and more work is needed to check whether our data can exclude this
possibility.

To summarize, we have presented clear evidence that a spin gap opens in the
quasi-1D vanadium oxide VOSb$_{2}$O$_{4}$ from susceptibility,
magnetization and ESR data. The overall behaviour strongly suggests that
this is due to the inherent SP instability of this spin 1/2 chain, but with
very strong fluctuations. Given the lack of inorganic materials exhibiting
this physics so far, the properties of this system are likely to attract a
lot of attention in the future.

\end{document}